\documentclass[journal=acnano,manuscript=article,floatfix]{achemso}
\usepackage{lipsum}
\usepackage{graphicx}
\usepackage{amsmath}
\usepackage{amssymb}
\usepackage{epsfig}
\usepackage{subeqnarray}
\usepackage{color}
\usepackage{bbold}
\usepackage{dsfont}
\usepackage{tikz}
\usepackage{cancel}
\usepackage{float}
\usetikzlibrary{positioning,calc}
\usepackage[version=3]{mhchem}
\usepackage{color}
\usepackage[normalem]{ulem}
\usepackage{float}

\newcommand{\fig}[1]{Figure~\ref{#1}}
\newcommand{\eq}[1]{Eq.~\ref{#1}}

\newcommand{\degree}{\ensuremath{^\circ}}

\newcommand{\ie}{\textit{i.e.},~}
\newcommand{\via}{\textit{via~}}
\newcommand{\eg}{\textit{e.g.,~}}

\newcommand{\albn}{\ce{Al_{0.93}B_{0.07}N} }  
\newcommand{\albns}{\ce{Al_{0.93}B_{0.07}N}}

\graphicspath{{./Figures/}}

\author{Walter J. Smith}
\affiliation{School of Mechanical Engineering and Birck Nanotechnology Center, Purdue University, West Lafayette, 47907, IN, USA}
\author{Betul Akkopru-Akgun}
\author{Erdem Ozdemir}
\affiliation{Department of Materials Science and Engineering and Materials Research Institute, The Pennsylvania State University, University Park, PA 16802, USA}
\author{Bogdan Dryzhakov}
\affiliation{Center for Nanophase Materials Sciences, Oak Ridge National Laboratory, Oak Ridge, TN, USA}
\author{John Hayden}
\author{Jon-Paul Maria}
\affiliation{Department of Materials Science and Engineering and Materials Research Institute, The Pennsylvania State University, University Park, PA 16802, USA}
\author{Kyle P. Kelley}
\affiliation{Center for Nanophase Materials Sciences, Oak Ridge National Laboratory, Oak Ridge, TN, USA}
\author{Clive A. Randall}
\author{Susan Trolier-McKinstry}
\affiliation{Department of Materials Science and Engineering and Materials Research Institute, The Pennsylvania State University, University Park, PA 16802, USA}
\author{Thomas E. Beechem}
\email{tbeechem@purdue.edu}
\affiliation{School of Mechanical Engineering and Birck Nanotechnology Center, Purdue University, West Lafayette, 47907, IN, USA}

\title{Nitrogen Vacancies Induce Fatigue in Ferroelectric \albn}

\begin{document}

\begin{abstract}
Wurtzite ferroelectrics (\eg \albns) are being explored for high-temperature and emerging near-, or in-compute, memory architectures due to the material advantages offered by their large remanent polarization and robust chemical stability. Despite these advantages, current \albn devices do not have sufficient endurance lifetime to meet roadmap targets. To identify the defects responsible for this limited endurance, a combination of electronic measurements and optical spectroscopies characterized the evolution of defect states within \albn with cycling. Ultrathin ($\sim$10 nm) metal contacts were used to optically probe regions subject to ferroelectric switching; photoluminescence spectroscopy identified the emergence of a transition near 2.1 eV whose intensity scaled with the non-switching polarization quantified \via positive-up negative-down (PUND) measurements. Accompanying thermally stimulated depolarization current (TSDC) and modulus spectroscopy measurements also observed the strengthening of a state near 2.1 eV. The origin of this feature is ascribed to transitions between a nitrogen vacancy and another defect deeper in the bandgap. Recognizing that the impurity concentration is largely fixed, strengthening of this transition indicates an increase in the number of nitrogen vacancies. Switching, therefore, creates vacancies in \albn likely due to hot-atom damage induced by the aggressive fields necessary to switch wurtzite materials that ultimately limits endurance. 
\end{abstract}


\newpage
Ferroelectrics show promise for next-generation non-volatile memory devices due to their energy efficiency, scalability, and potential for short access times. \cite{zhu_2023} Wurtzite ferroelectrics, in particular, offer differentiating advantages arising from their high spontaneous polarization, retention, and capability for high-temperature operation.\cite{fichtner_2019,zhu_2021,zhu_2023,kim_2023a,drury_2022,wang_2023d} Boron-doped AlN (\albns) exemplifies many of these advantages, exhibiting a remanent polarization of $\sim$ 125 \ce{\mu C}/\ce{cm^2},\cite{hayden_2021} good scaling of the polarization down to 20 nm, \cite{casamento_2024} and excellent retention.\cite{zhu_2023} Owing to these factors, as well as its 
complementary metal-oxide semiconductor (CMOS) compatibility, \albn is being pursued for in-memory compute applications, high-temperature memory elements, as well as negative drain induced barrier lowering at semiconductor interfaces.\cite{casamento_2024}

\albn exhibits limited cycling endurance, however.  Widespread adoption as a non-volatile memory element typically necessitates reliability of $10^{12}$ read and write cycles, \cite{ieee_2022c} yet \albn devices typically show substantive leakage-induced degradation by $\sim$\ce{10^6} cycles before hard dielectric breakdown.\cite{he_2024} The limited endurance is correlated to the high coercive field ($E_c$) to breakdown field ($E_{BD}$) ratio ($E_c/E_{BD} \approx 80 \%$ for \albns).\cite{zhu_2021} High-fields, in turn, are more likely to induce defect formation associated with the atomic movement implicit to switching in a wurtzite ferroelectric through a process termed ``hot-atom" damage.\cite{masuduzzaman_2014} Point defects, in turn, lead to leakage currents; the associated Joule-heating accelerates the formation of additional defects.\cite{guido_2023}  Ultimately, the number of defects reach a threshold sufficient to induce thermal breakdown.

Quantifying the evolution of point defects in \albn is therefore necessary to identify methods for enhancing endurance.  In this work, the evolution of defects in \albn is studied as they are electrically cycled using a combination of optical and electrical characterization. Specifically,  photoluminescence (PL), cathodoluminescence (CL),  and impedance spectroscopies along with thermally stimulated depolarization current (TSDC) measurements together identify an increase in defects related to the formation of nitrogen vacancies that emerge with cycling and fatigue in \albn capacitors. Taken in total, this study underscores the implicit relationship of defects to the switching behavior, lifetime, and performance of wurtzite ferroelectric devices.


\section{Results and Discussion}

Thin films of \albn ($\sim$ 190 nm) were sputtered on tungsten bottom electrodes as described in a previous report.\cite{hayden_2021} The impurity level in these films is believed to be very low, albeit with a non-zero concentration of oxygen. 
The metal-ferroelectric-metal (MFM) stack was fabricated by subsequent photolithography and sputter deposition of an additional 5 nm of  W then 5 nm of Pt without breaking vacuum, which serves as the top contact. This ultra-thin, semi-transparent, metal contact enables photoluminescence (PL) and cathodoluminescence (CL) measurements through the contact, allowing for optical characterization of devices subject to field-cycling. For complementary electrical measurements, the sample stack was completed \textit{ex situ} by photolithography and sputter deposition of an additional 100 nm of Pt. 

To assess changes in the \albn with switching, devices were cycled by applying bipolar triangle waveforms at an electric field of 6 MV/cm at 400 Hz. A subset of measurements utilized a unipolar triangular waveform.  Unless otherwise specified, cycling in the text refers to bipolar waveforms causing ferroelectric switching between positive and negative polarization.  

At regular intervals during the cycling, PUND (positive up negative down) measurements were performed. The waveforms for these measurements, along with those utilized for ferroelectric switching, are shown schematically in the Supporting Information (see Figure S1). Many of the ultra-thin contacts eroded at their perimeters throughout their lifetime, however, as a consequence of field concentration at the contact edge that causes localized breakdown.\cite{he_2024} Despite the erosion, the devices continue to operate albeit with a smaller effective area (see Figure S2 in Supporting Information).  To account for this smaller area, reported polarization values were calculated using optical micrography to identify, and account for, those regions still covered by the contact metal at a given number of switching cycles. Complementing these PUND measurements, modulus, thermally stimulated depolarization current (TSDC), photoluminescence and cathodoluminescence data were collected to identify both the origin and energy levels of the defects that emerge with device fatigue. Details surrounding the implementation of these techniques are provided in the Methods Section. 



\begin{figure}
    \centering
    \includegraphics[width=\textwidth]{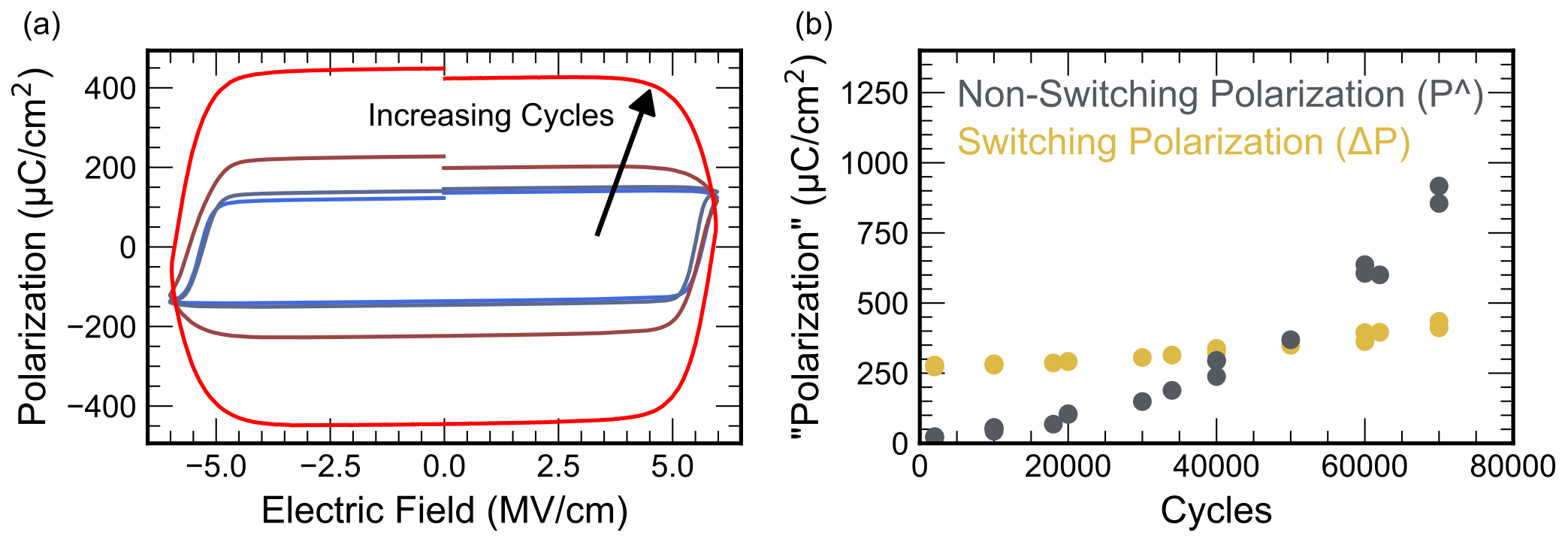}
    \caption{(a) Change in hysteresis of the apparent polarization in \albn with switching cycles. Although the behavior is nominally constant for the first $\mathrm{10^3}$ cycles, an increase in the apparent remanent polarization, rounding of the tips of the loop, and an ``inflation" of the hysteresis curves is indicative of increases in leakage current.  (b) This is corroborated by PUND measurements where the non-switchable polarization increases with cycling.}
    \label{Fig_1}
\end{figure}

\albn devices exhibited square hysteresis curves with remanent polarizations of 125 \ce{\mu C/cm^2} and coercive fields of 5.5 MV/cm, as shown \fig{Fig_1}.  These values are typical of \albns.\cite{hayden_2021}  With cycling, PUND measurements show a switching polarization ($\Delta P$) that stays comparatively constant. The slight increase in $\Delta P$ is associated with incomplete compensation of the increased leakage and/or small errors accounting for changes in electrode area. In contrast, the non-switchable polarization ($P^{\wedge}$) increases significantly. This increase is accompanied by an ``inflation" of the hysteresis curves that together indicates a rise in leakage with cycling. The following focuses on understanding the physical changes in the material that are the origin for the cycling-induced leakage.  

Photoluminescence (PL) spectroscopy uses incident photons to excite electrons into higher energy states that can induce photon emission as the electrons relax to lower energy states. Using sub-bandgap laser excitation (405 nm, 3.06 eV), direct absorption does not occur across the $\sim$5.8 eV bandgap of \albns.\cite{hayden_2021} Photoluminescence, therefore, preferentially probes electronic transitions involving defect states within the \albn bandgap. 

\begin{figure}
    \centering
    \includegraphics[width=\textwidth]{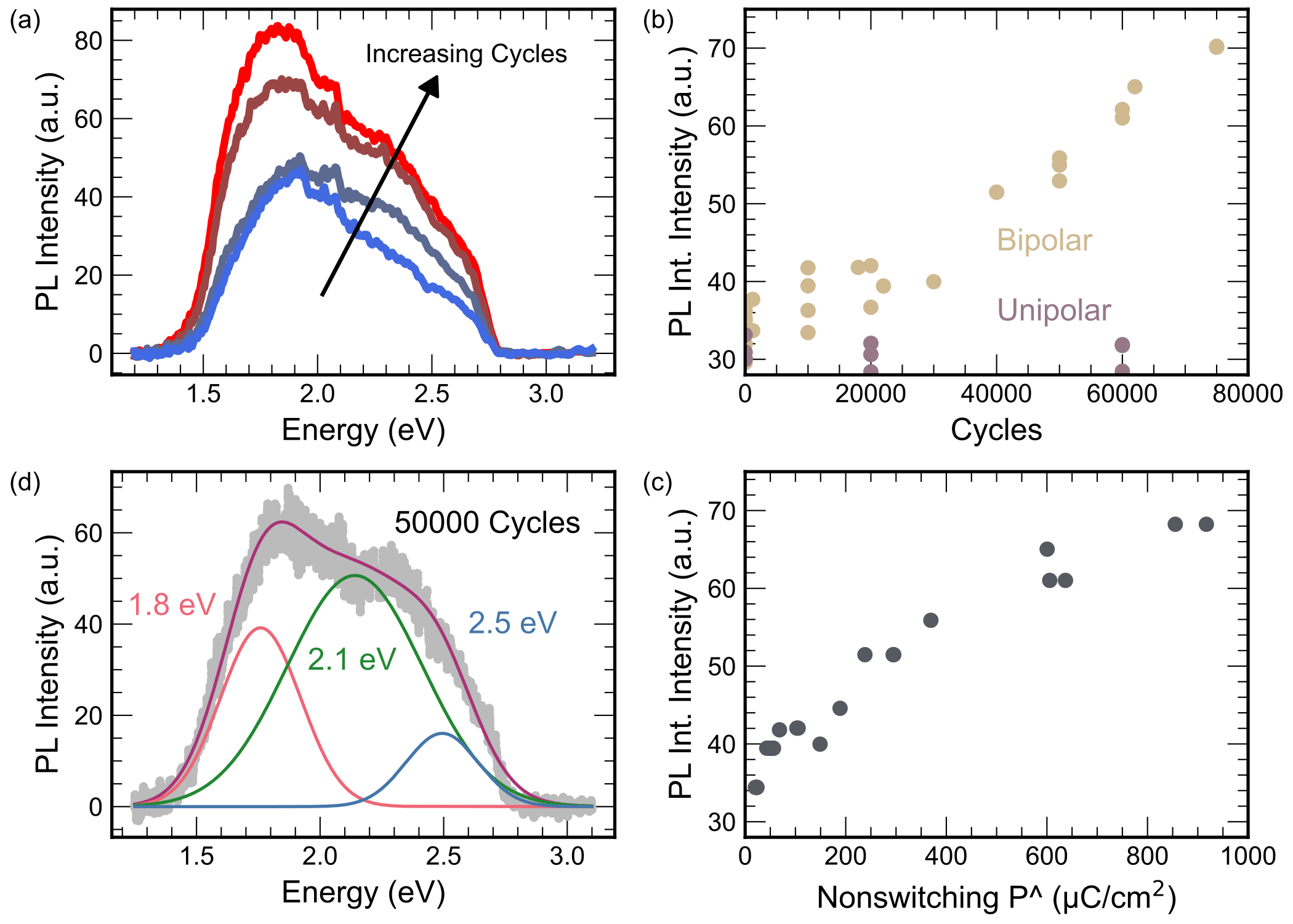} 
    \caption{(a) Change in photoluminescence (PL) spectra with cycling. (b) The total integrated photoluminescence intensity increases and is accompanied by a change in spectral shape with bipolar switching. Intensity does not change when the device is subject to unipolar switching.  (c)  These changes of PL-intensity correlate strongly with non-switching polarization.(d) Raw data and accompanying spectral fit made up of 3 Gaussian lineshapes for a device subjected to 50,000 cycles.}
    \label{Fig_2}
\end{figure}

Cycling in \albn causes distinct, repeatable, and significant changes in the PL-signal [see \fig{Fig_2}(a)]. Total intensity increases in a manner that is positively correlated with the change in the non-switchable polarization [see \fig{Fig_2}(b,c)].  Switching, rather than just the application of an electric field, drives these changes, as the PL-intensity remained nominally constant when capacitors were subjected to a unipolar field.  To quantify this cycle-dependent variation, PL-spectra were fit to the sum of three Gaussian features that identified transitions having energies of 1.76$\; \pm \;$0.05 eV, 2.10$\; \pm \;$0.06 eV, and 2.51$\; \pm \;$0.05 eV, as shown by the representative fit in \fig{Fig_2}(d). The positions of these features did not appreciably vary with switching. Fitting, in turn, enables monitoring of each feature's relative intensity with cycling while also providing a means for comparison with electrical measurements and theoretically predicted defect energy levels.  

\begin{figure}
    \centering
    \includegraphics[width=\textwidth]{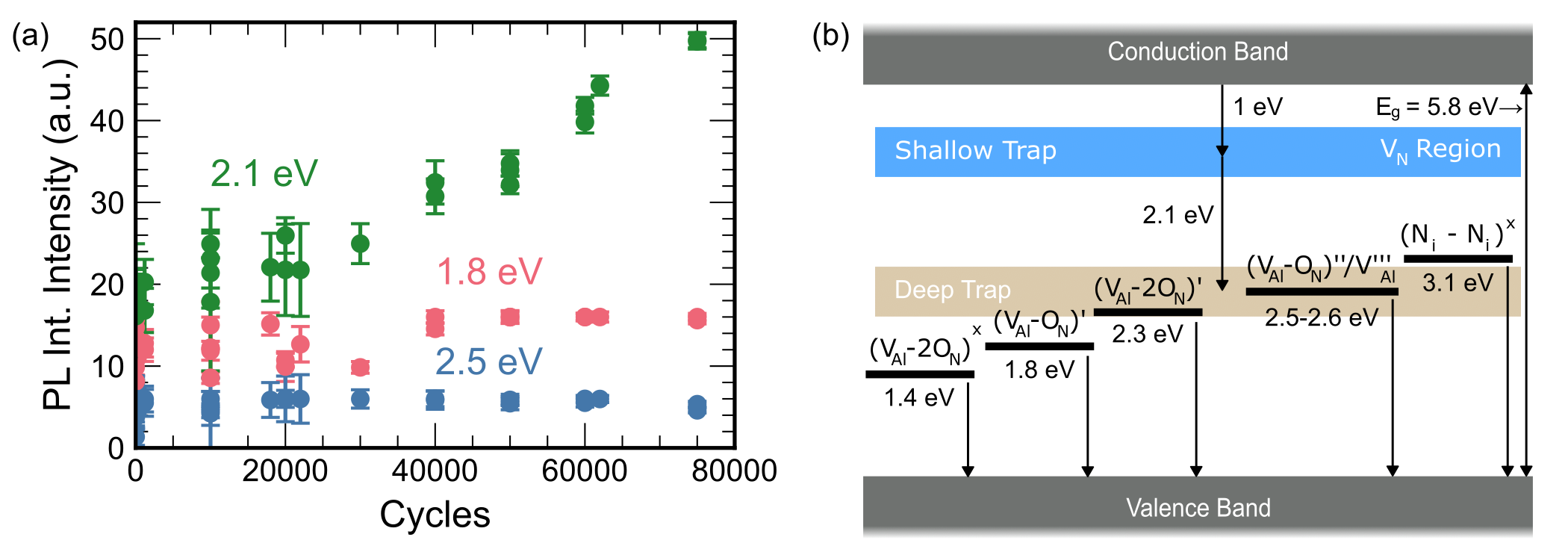} 
    \caption{(a) PL-integrated intensity dependence versus cycle for each of the three identified transitions.  Only the 2.10 eV peak significantly changes with cycling. (b) Estimated energies of defect states in the bandgap of \albns, using AlN as a reference.\cite{zhou_2020, koppe_2016,aleksandrov_2020a} Nominal regions of shallow to deep defect-to-defect transitions are denoted by boxes.}
    \label{Fig_3}
\end{figure}

The cycle-dependent increase in PL-intensity is primarily due to the growing influence of the 2.1 eV transition. As shown in \fig{Fig_3}(a), intensities of both the 1.76 and 2.51 eV features remain nearly constant while the response centered at 2.1 eV more than doubles with switching. Recognizing that the increase in total PL-intensity correlates with the non-switching polarization [see \fig{Fig_2}(c)], the dependence of just the 2.1 eV feature on cycling implies that a transition having this energy is in some way correlated with the degradation in ferroelectric performance. Sub-bandgap excitation, in turn, dictates that at least one of the states involved in the 2.1 eV PL-feature comes from a defect.  This is further supported by the general acceptance that many luminescent features in the range of 2.1-2.3 eV observed in AlN are due to transitions between shallow donor and deep acceptor defects (\ie defect-to-defect).\cite{aleksandrov_2020,slack_2002,sarua_2003} Together, this indirectly suggests that changes in a particular, or small set, of defect(s) giving rise to this 2.1 eV transition accompany the increase in non-switchable polarization and leakage current.

The following seeks to specify the origin of the emerging defects correlated with this transition near 2.1 eV that accompany reduced ferroelectric performance.  Based on the low-impurity concentrations of our films and the tendency for AlN-based compounds to be n-type,\cite{vandewalle_2003b,yan_2014a} calculations of formation energy indicate that the most likely defects are those originating from: intrinsic vacancies (\eg $\mathrm{V_{Al}^{'',(''')}}$, $\mathrm{V_N^{\bullet, (\bullet \bullet), (\bullet\bullet\bullet)}}$), nitrogen split-interstitials [$\mathrm{(N_i - N_i)^{x,(')}}$] and complexes associated with the incorporation of oxygen into the lattice.\cite{aleksandrov_2020a,yan_2014a}  Specifying the defect impacting performance is thus a task in assigning a particular defect from these candidates to the 2.1 eV transition energy. At the time of this writing, however, literature on defect state energies particular to \albn are not reported. All subsequent analysis is, therefore, based on comparison to the sister binary compound AlN instead.  It is assumed that defect energies remain at similar relative locations within the bandgap of \albn as they do in AlN, albeit scaled to the smaller bandgap of the alloy.  

Oxygen impurity defects and their associated complexes are not the primary source of the \albn degradation.  This is deduced \via the insensitivity of modes having peak energies of $\sim$1.8 eV seen in both photo- and cathodoluminescence along with a complementary cycle-insensitive mode observed at 3.1 eV in CL.  The PL spectrum of AlN in the visible portion of the spectrum is known to be dominated by radiative recombination events involving metal vacancies and oxygen impurities \textemdash $\mathrm{(V_{Al} - O_N)^{',('')}}$ and $\mathrm{(V_{Al} - 2O_N)^{',(x)}}$, termed here an O-complex. \cite{zhou_2020,koppe_2016,aleksandrov_2020a,yan_2014a} A PL-feature near 1.9 eV, for example, has been consistently linked to the $\mathrm{(V_{Al} - O_N)^{',('')}}$ to VBM transition in AlN.\cite{sedhain_2012,zhou_2020}  Recognizing that the bandgap of \albn (5.8 eV) is $\approx \mathrm{95}\%$ of AlN (6.1 eV), the observed mode at 1.76 eV (\ie 1.76/1.90 = 0.93) is, therefore, assigned to this O-complex to VBM transition [see \fig{Fig_3}(b)]. 
\begin{figure}[t]
    \centering
    \includegraphics[width=\textwidth]{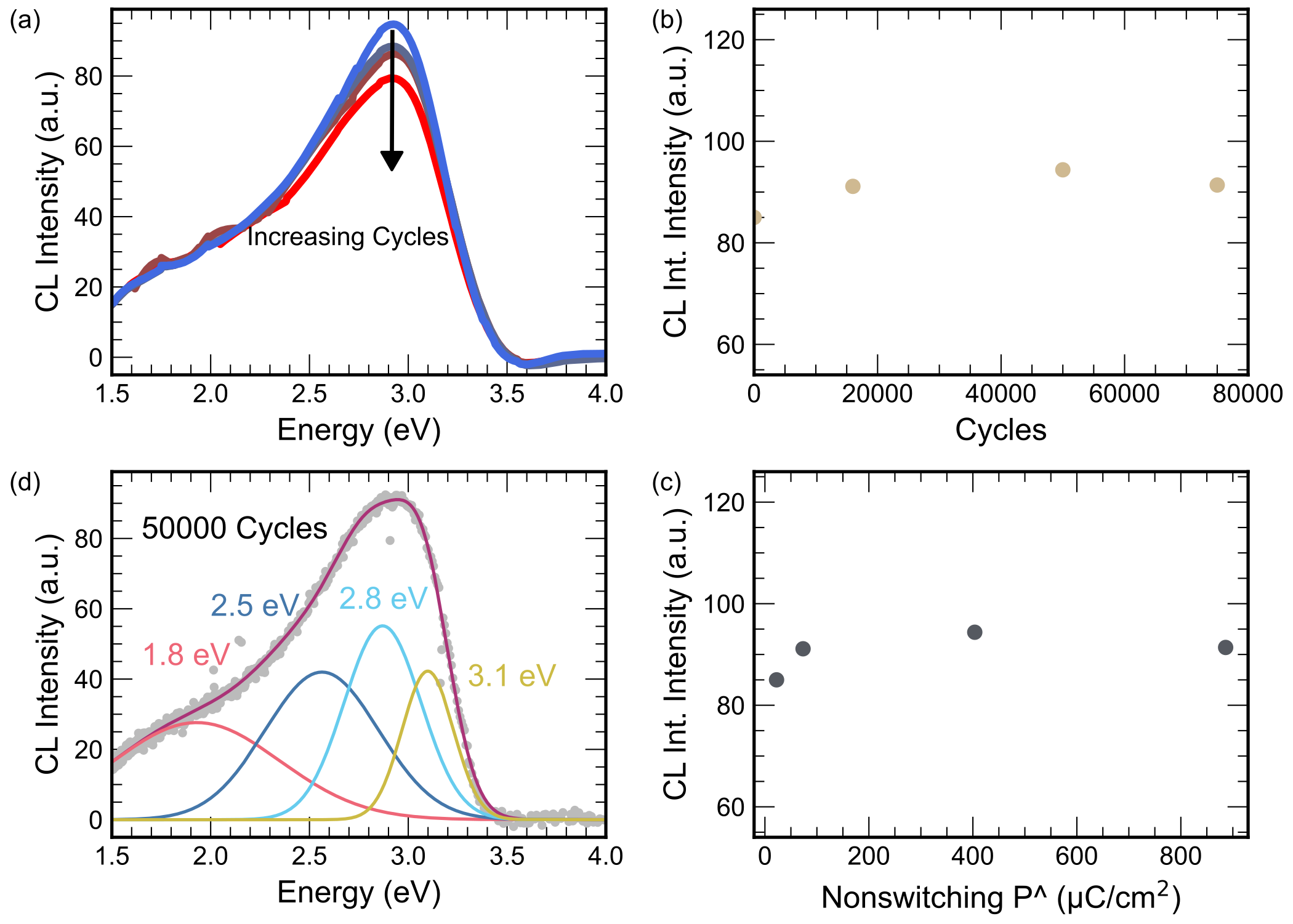} 
    \caption{(a) Change in cathodoluminescence (CL) spectra with cycling. Unlike that observed in PL, (b) the integrated CL intensity remains relatively constant as the device is cycled and, therefore, shows little correlation with (c) non-switchable polarization. (d) Raw data and accompanying spectral fit made up of 4 Gaussian lineshapes for a device subject to 50,000 cycles.}
    \label{Fig_4}
\end{figure}

CL-spectra support this deduction by observing the same mode near 1.8 eV, its complementary transition from the CBM to this O-complex (3.1 eV), and a near constant intensity with ferroelectric switching. \fig{Fig_4}(a), for example, shows the relatively small changes in CL-spectra that occur with ferroelectric switching.  Unlike the PL-spectrum whose intensity more than doubles,  the total CL integrated intensity changes by only $8\%$ [compare \fig{Fig_2}(b) and \fig{Fig_4}(b)].  Fitting of the CL-spectra identifies features at 1.8, 2.5, 2.8, and 3.1 eV.  Of note, inclusion of a mode at 2.1 eV\textemdash like that observed to vary greatly in PL\textemdash generated unreasonable profiles. It is, therefore, concluded that CL was insensitive to the transition correlated to the cycle-dependent changes in ferroelectric performance.  The cause for CL's insensitivity to this transition will be addressed in more detail later. 

Having an energy nominally equal to that observed in PL, the 1.8 eV transition seen in CL is deduced to arise from the same O-complex to VBM transition.  The 3.1 eV mode, meanwhile, is likely a complementary CBM to O-complex transition. Such complementary pairs indicate transitions of a defect state with both the CBM and VBM.  This deduction follows previous reports where pairs of spectral features together exhibit energies spanning the entirety of the bandgap when considering lattice relaxation.\cite{koppe_2016,zhou_2020} For AlN, the relaxation energy is $E_R\approx$0.4 eV for transitions involving O-complexes.\cite{yan_2014a} Recognizing that this relaxation will occur for both the transition from the O-complex to VBM ($E_{(O\text{-}X) \rightarrow VBM}$=1.8 eV) and CBM ($E_{CBM\rightarrow(O\text{-}X)}$=3.1 eV), the total energy associated with both transitions compares well with the 5.8 eV bandgap of \albn [$(E_{(O\text{-}X) \rightarrow VBM}+E_R) + (E_{CBM\rightarrow(O\text{-}X)}+E_R)=\mathrm{5.7 \; eV}$]. Thus, since both the 1.8 and 3.1 eV responses are linked to an O-complex and the signals associated with them remain relatively constant with cycling in both PL and/or CL, it is concluded that these defects are not playing a primary role in the ferroelectric performance reduction with switching. This deduction is consistent with recent reports in AlScN where oxygen impurities were shown to possibly improve the endurance of wurtzite-based ferroelectric devices.\cite{islam_2025}  

Applying similar logic to that used in analyzing the O-complexes, single aluminum vacancies ($\mathrm{V_{Al}^{'',(''')}}$) are not the primary cause of the cycle dependent performance degradation either.  Experimental\cite{sedhain_2012,zhou_2020} and theoretical\cite{yan_2014a,aleksandrov_2020a} predictions, for example, each assign a strong emission near 2.75 eV to transitions between aluminum vacancies ($\mathrm{V_{Al}^{'',(''')}}$) and the material's intrinsic bands (VBM, CBM). When scaled to the bandgap of \albn, this corresponds to a transition near 2.5 eV observed to remain constant with cycling in both PL [\fig{Fig_3}(a)] and CL [\fig{Fig_4}(b)]. The associated mode at 2.8 eV observed to remain constant in CL, meanwhile, is likely a complementary transition from the $\mathrm{V_{Al}^{'',(''')}}$ to the other intrinsic band (VBM or CBM), where the predicted lattice relaxation energy of $\approx$0.3 eV associated with transitions of aluminum vacancies in AlN has been used in this deduction.\cite{yan_2014a,aleksandrov_2020}  Thus, it is concluded that aluminum vacancies are not the primary source of the cycle dependent degradation based upon their constant response in PL and CL-spectra. 

As alluded to previously, transitions in the luminescent response of AlN in the 2-2.3 eV range are typically linked to defect-to-defect transitions between shallow electron trap states and deeper acceptors. Previous reports have linked the shallow traps giving rise to these signals to impurities of carbon,\cite{aleksandrov_2020} silicon,\cite{lamprecht_2017a} isolated oxygen,\cite{sarua_2003} and nitrogen vacancies,\cite{aleksandrov_2020a} for example.  As it relates to \albn films studied here, impurities beyond oxygen are largely absent.  The oxygen in the film, meanwhile, will preferentially form a complex with aluminum vacancies at the low-levels at which it is present. It is unlikely to exist in its isolated form.\cite{maki_2011} Therefore, nitrogen vacancies become a likely candidate for the shallow donor giving rise to the observed 2.1 eV transition, especially considering their low formation energy.\cite{aleksandrov_2020a} As this transition grows stronger with cycling while deeper O-complex and aluminum vacancy states remain constant, this implies that nitrogen vacancies\textemdash and possibly complementary nitrogen interstitial states\textemdash are formed during the switching event. 

Implied within this conclusion, in turn, is that the increase in nitrogen vacancy concentration is coupled to the ultimate endurance of \albn devices.  The central link between nitrogen vacancies and endurance within \albn devices is the primary outcome of this study.  The following supports this finding through TSDC measurements and modulus spectroscopy.

\begin{figure}
    \centering
    \includegraphics[width=\textwidth]{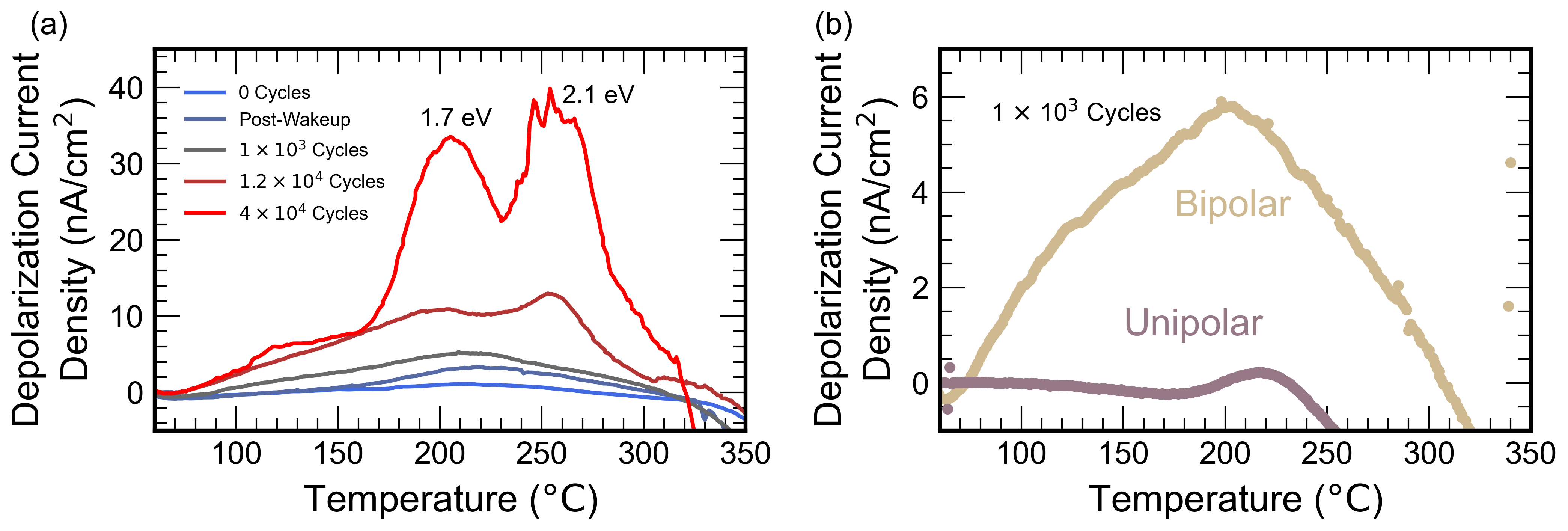} 
    \caption{(a) TSDC data of \albn films degraded at 150$\degree$C, under 200 kV/cm, for 30 minutes for varying cycles, showing the increase in response believed to arise from $\mathrm{V_{N}^{\cdot \cdot \cdot}}$ and $\mathrm{(N_i - N_i)}$ related transitions at 1.7 eV and 2.1 eV. (b) TSDC of \albn  following bipolar switching and non-switching unipolar excitation; a much more significant defect concentration is apparent following bipolar cycling.}
    \label{Fig_5}
\end{figure}

TSDC, for instance, also observes an increasing prevalence of a transition near 2.1 eV with cycling. This is seen qualitatively in the continual increase in depolarization current that occurs with cycling, as shown in \fig{Fig_5}(a), indicating the generation of defects. As with PL, the TSDC peaks do not, however, increase under unipolar cycling [see \fig{Fig_5}(b)]. Along with this general increase in the depolarization current, two prominent peaks emerge in the TSDC data at 204$\degree$C and 258$\degree$C for samples exposed to more than 1,000 bipolar cycles. Using the full-width half-maximum method, the activation energies of these peaks are found to be 1.7$\; \pm \;$0.1 eV and 2.1$\; \pm \;$0.2 eV.  The 2.1 eV activation energy, as described above, is consistent with the transition between a shallow donor\textemdash likely a nitrogen vacancy\textemdash and a deep acceptor.  
\vspace{\baselineskip}

Because the 1.7 eV feature also grows with cycling, it is considered a distinct process from the switching insensitive 1.8 eV mode observed in both PL and CL that was attributed to an O-complex to VBM transition.  The 1.7 eV feature measured in TSDC is instead believed to arise from an additional shallow donor to deep acceptor transition. Responses in this lower energy range have been linked to a variety of shallow impurity\textemdash including those related to nitrogen vacancies\textemdash to deep oxygen-related defects in several photo- and thermoluminescence reports .\cite{aleksandrov_2020,aleksandrov_2015,lamprecht_2017,lamprecht_2017a} Being a defect-to-defect transition explains why both PL and CL were ``blind" to the emergence of this 1.7 eV mode with cycling since it was competing with the near degenerate defect-to-band transition at 1.8 eV. Defect-to-band transitions are typically much stronger in their optical response and tend to overwhelm defect-to-defect transitions owing to the much larger density of states corresponding to the material's intrinsic bands. This same argument explains why CL does not see the growing 2.1 eV mode either since it preferentially excites carriers over the bandgap and thus is dominated by band-to-defect transitions. Finally, it is noteworthy, albeit speculative, that the energy difference between the split-interstitial nitrogen defect [\ie $\mathrm{(N_i - N_i)^{x,(')}}$] and $\mathrm{V_{N}^{\bullet \bullet \bullet}}$ state are predicted to differ by 1.8 eV, which would imply that ferroelectric switching displaces some nitrogen atoms from their place in the lattice creating both vacancies and interstitial defects.\cite{aleksandrov_2020a}  Taken in aggregate, these TSDC results support the view that the nitrogen vacancy concentration increases with ferroelectric switching. This, in turn, increases the leakage current.

\begin{figure}
    \centering
    \includegraphics[width=80mm]{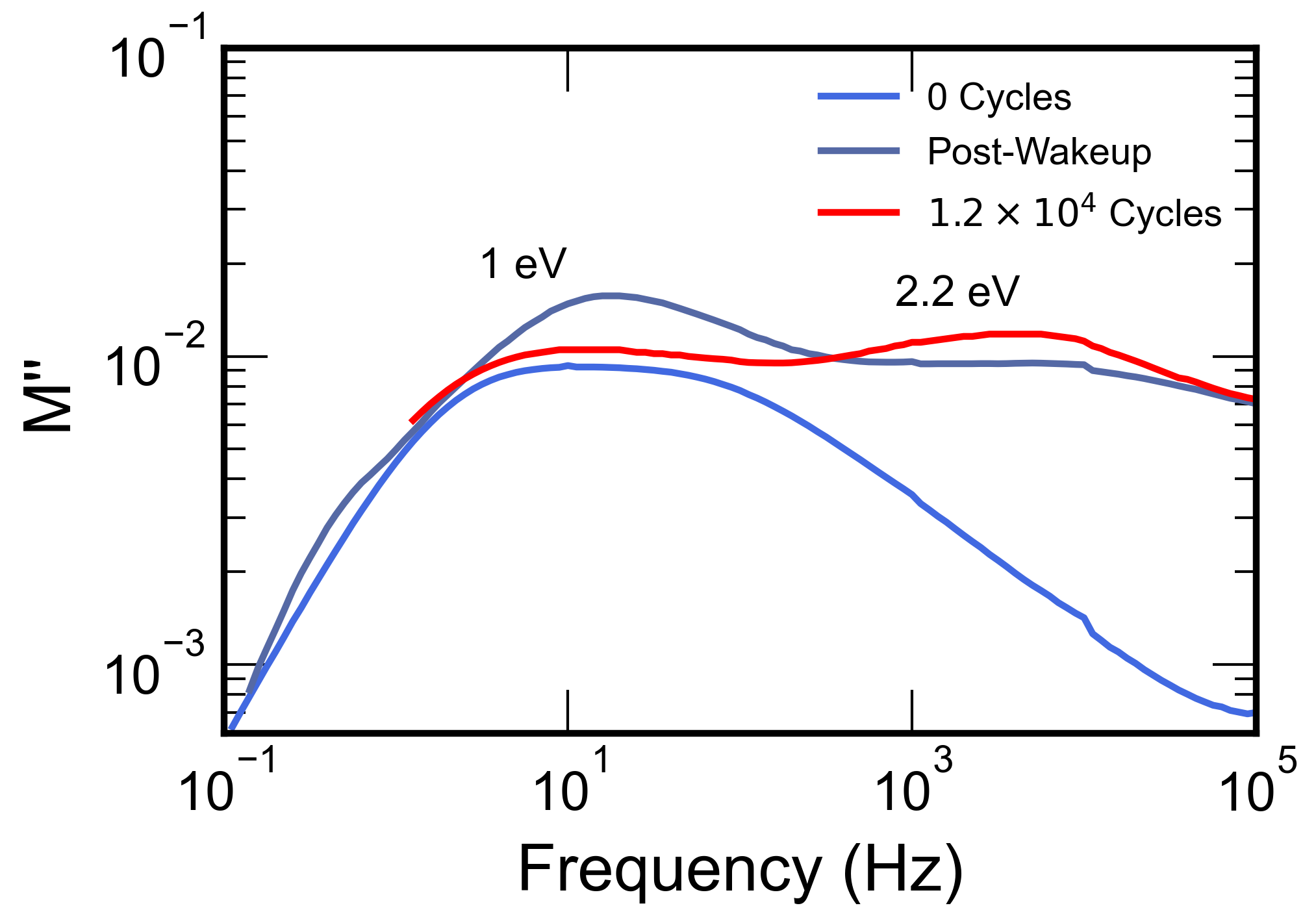} 
    \caption{Modulus spectroscopy of pristine and fatigued (12,000 cycles) films are measured at 380$\degree$C using a 100 mV alternating current (AC) signal from 1 MHz to 100 mHz. Two conduction pathways at $\sim$ 1 eV and 2.2 eV are defined. As the \albn films are cycled, the 1 eV peak does not increase appreciably. The defect at 2.2 eV increases as the \albn is cycled.}
    \label{Fig_6}
\end{figure}

Modulus spectroscopy further corroborates an increase in electrical conductivity associated with nitrogen-vacancy related defects with cycling. The modulus spectra in \fig{Fig_6} show two broad maxima indicative of inhomogeneous bulk conductivity, potentially due to an inhomogeneous distribution of defects throughout the film. The higher frequency peak ($\sim$\ce{10^3} Hz) corresponds to higher conductivity [\textit{c.f.} Eq. \ref{Eqn_ModulusSpectroscopy} ].  With cycling, the high-frequency maximum becomes more pronounced, suggesting an increase in conductivity of the film with bipolar switching similar to the observed enhancement in non-switching polarization seen in \fig{Fig_1}(b). 

\begin{figure}[htbp]
    \centering
    \includegraphics[width=\textwidth]{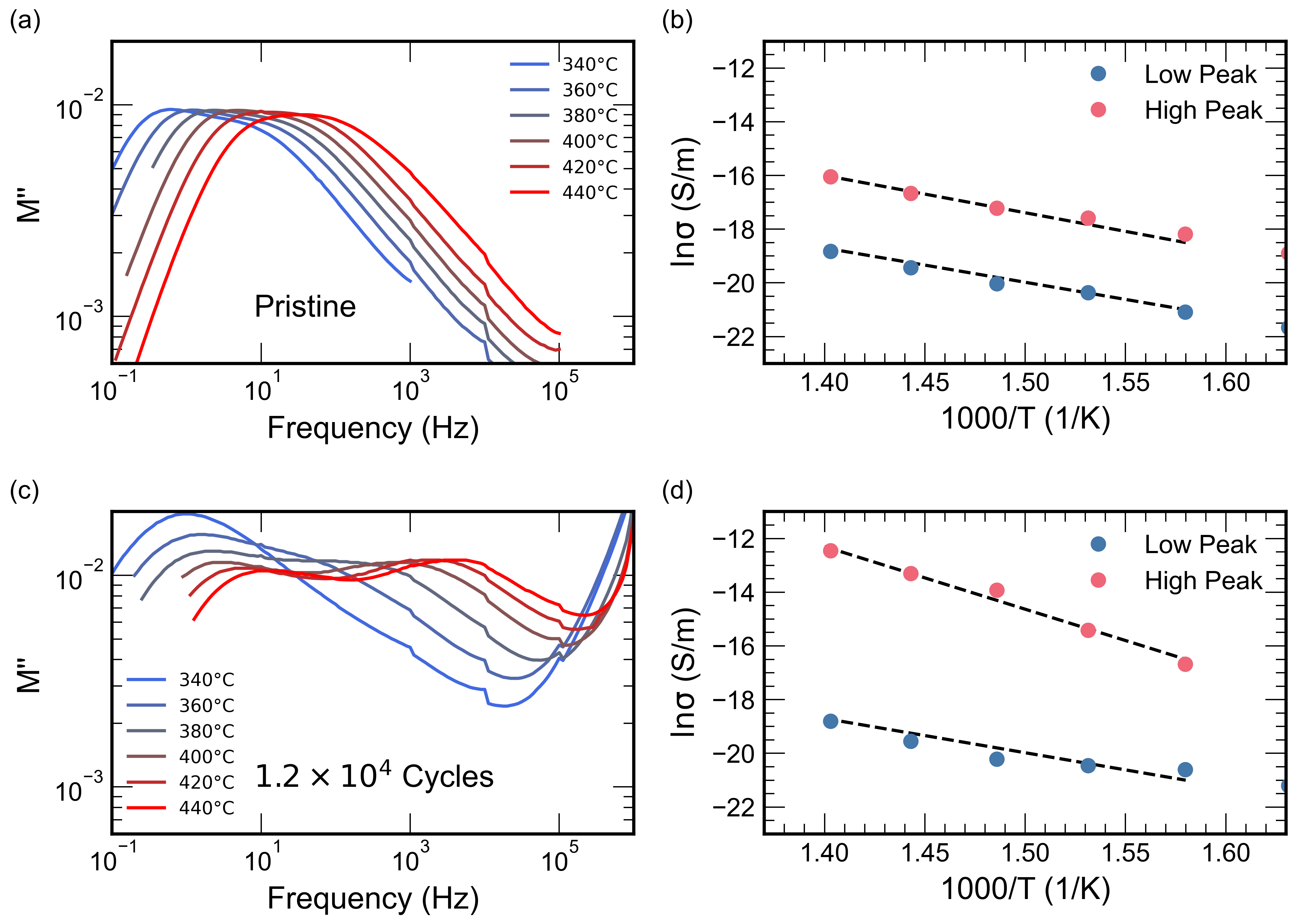} 
    \caption{Temperature dependent modulus spectroscopy data for (a) pristine and (c) fatigued films at temperatures ranging from 340$\degree$ to 440$\degree$C. By fitting maxima in the modulus spectra as a function of temperature, slopes identified activation energies dictating transport within the (b) pristine and (d) fatigued films.}    
    \label{Fig_7}
\end{figure}

The activation energies associated with the low and high frequency maxima of the modulus were determined from their temperature dependence, as shown in \fig{Fig_7}. From this analysis, the low frequency mode exhibited an activation energy near 1 eV in both the pristine and cycled films. This transition energy is consistent with thermal activation of electrons from shallow nitrogen vacancy traps into the conduction band.\cite{sedhain_2012,yan_2014a,he_2024} It also provides further evidence for the assignment of nitrogen vacancies as the the shallow trap in the shallow to deep defect-to-defect transitions observed in PL and TSDC. Thus, the low frequency modulus peak is attributed to bulk conduction mediated by nitrogen vacancies. 

The high frequency modulus peak, in contrast, changes significantly with cycling. In its pristine state, a similar activation energy of 1.04 $\pm$ 0.08 eV was found, suggesting a region with a higher concentration of nitrogen vacancies than the rest of the film. For fatigued films, however, the high frequency peak has a substantially higher activation energy of 2.2 $\pm$ 0.29 eV. This energy is comparable to that observed in the cycling PL and TSDC data and, therefore, attributed to the same nitrogen-vacancy related, shallow to deep, defect-to-defect transitions ascribed previously. 

As both features in the modulus spectra shift to higher frequency with cycling, the data suggests that the defect(s) associated with these features become increasingly active with cycling. This could occur because of hot-atom damage wherein the movement of atoms during ferroelectric switching induces the generation of defects. Wurtzite ferroelectrics are predisposed to hot-atom damage because the effect becomes stronger as switching fields approach that of breakdown, as in \albns.\cite{masuduzzaman_2014,guido_2023b} Importantly, these data also suggest a reduced concentration of nitrogen vacancies should lower the baseline conductivity in the samples and may delay the onset of cycling-induced degradation. This, in turn, may be helpful in increasing the fatigue lifetime of wurtzite films.

\section{Conclusions}
Nitrogen vacancies are central to the cycling endurance of ferroelectric \albns.  This conclusion is reached based on the correlation between the increase in non-switching polarization and the emergence of an electronic transition having an energy separation of 2.1 eV observed in photoluminescence, TSDC, and modulus measurements.  By using sub-bandgap excitation energies when performing PL in conjunction with CL and electronic characterization, the emergent transition was identified as charge moving from a nitrogen vacancy to another defect within the bandgap of \albns. The increased prominence of this conductive pathway with cycling, in turn, indicates that fatigue in \albn devices arises from the formation of nitrogen vacancies that occurs with ferroelectric switching.  
\section{Methods}

\subsubsection{Photoluminescence (PL) Measurements}
Photoluminescence spectra were acquired on a WITec Alpha 300R spectral imaging system with a 405 nm (3.05 eV) exciting laser irradiating the device at 3.0 mW across a diffraction limited area realized using a 50X/0.55 NA objective. The spectra were gathered using a UHTS 600 VIS spectrometer with a spectral resolution of less than 0.05 eV. A custom-built probe station beneath the spectral imaging system allowed ferroelectric measurements and optical testing without having to re-contact devices, or re-focus the system, between tests.

\subsubsection{Cathodoluminescence (CL) Measurements}
Cathodoluminescence (CL) spectra were collected using an FEI Quattro environmental scanning electron microscope (SEM) integrated with a Delmic Sparc CL detection module. This configuration simultaneously records secondary electron images and CL emission. An optical filter ensured that only CL-signals with energy less than 3.35 eV were collected. The measurements were conducted at an electron accelerating voltage of 10 kV and a beam current of 740 pA. Data was acquired and averaged over a 49 µm\ce{^2} scan area with 100 separate spectra collected evenly over the area (\ie 10 x 10 pixels).

\subsubsection{Thermally Stimulated Depolarization Current (TSDC) Measurements}
TSDC data were acquired on both pristine and fatigued \albn films using a HP 4140 pA meter to explore the evolution of defect dipoles, trapped charge, and space charge as a function of ferroelectric cycling. The TSDC experiment itself began by poling pristine films at an applied field of 0.2 MV/cm for 8 hours at 150$\degree$C followed by cooling to 50$\degree$C while the applied field was maintained. Following wakeup and extensive cycling, the samples did not survive such conditions and the poling time was reduced to 30 minutes. Subsequently, the electrodes across the \albn films were short-circuited and the depolarization current measured as the sample was reheated to 400$\degree$C under a constant heating rate of 10$\degree$C/min. Defect energies were extracted using the full-width half-maximum (FWHM) method in which the maxima in the depolarization current versus temperature data are fit, and then the FWHM of the feature is linked to the activation energy \via:\cite{chen_1981a} 
\begin{equation}
    W= \frac{2.30 k_B T_{max}^2}{\Delta T_{1/2}}
\end{equation}
where W is the activation energy, \ce{T_{max}} is the temperature where the current maximizes and $\Delta$T$_{1/2}$ is the FWHM of this feature. This activation energy quantifies the energy barriers associated with a given process and can, therefore, be used to assess transition energies involving defect states in a material.

\subsubsection{Modulus Spectroscopy Measurements}
Modulus spectroscopy was implemented using a Solatron Analytical Modulab. Both pristine and cycled films were heated to temperatures ranging from 340$\degree$C to 440$\degree$C whereupon impedance, $Z$, was measured at frequencies spanning from 100 mHz to 1 MHz under an applied AC field of 100 mV.  From the measured complex impedance, the complex modulus modulus, $M=M'-iM''$, is deduced \via
\begin{equation}
    M(\omega)=i\omega C_0Z(\omega)
    \label{Eqn_ImaginaryImpedence}
\end{equation}
where \ce{C_0} is the so called ``empty cell" capacitance and $\omega$ is the angular frequency. The resulting modulus relaxation frequency, $f_r$, is given by 
\begin{equation}
    f_r=\frac{\sigma}{2\pi \epsilon_0 \epsilon_r}
    \label{Eqn_ModulusSpectroscopy}
\end{equation}
where \ce{\sigma} is the bulk conductivity, \ce{\epsilon_0} is the permittivity of free space, and \ce{\epsilon_r} is the relative permittivity of the film. The imaginary component of the electric modulus will exhibit a maximum at the relaxation frequency given by \eq{Eqn_ModulusSpectroscopy} while shifts in this peak to higher frequency imply an increase in conductivity of the film.  The activation energies for conduction can be found from the slope of the \ce{ln(\sigma)} curves with respect to temperature.\cite{akkopru-akgun_2021} The analysis implicitly capitalizes upon the small temperature dependence of \albn permittivity.\cite{zhu_2021} 

\section{Acknowledgements}
Film synthesis, optical characterization, electrical characterization, and analysis were supported by the Center for 3D Ferroelectric Microelectronics (3DFeM), an Energy Frontier Research Center funded by the U.S. Department of Energy, Office of Science, Basic Energy Sciences under Award No. DE-SC0021118. Cathodoluminescence was supported by the Center for Nanophase Materials Sciences (CNMS), which is a US Department of Energy, Office of Science User Facility at Oak Ridge National Laboratory.

\section{Associated Content}
\subsubsection{Supporting Information Available:}
The Supporting Information is available free of charge at \textit{link-to-supporting-information}.

Waveforms applied to the \albn capacitors and calculation of switching polarization; and microscope images of field-breakdown of top contact.
\subsubsection{Data Availability}
The data that support the findings of this study are available from the corresponding author upon reasonable request.

\bibliography{Library_Updated}    

\end{document}